\documentclass[twocolumn,showpacs,amsmath,amssymb,superscriptaddress]{revtex4-1}
\usepackage{graphicx}% Include figure files
\usepackage{dcolumn}% Align table columns on decimal point
\usepackage{bm}% bold math
\begin{document}

%\preprint{}

\title{Characterizing the atomic mass surface beyond the proton drip line via $\alpha$-decay measurements of the $\pi s_{1/2}$ ground state of $^{165}$Re and the $\pi h_{11/2}$ isomer in $^{161}$Ta}
% Force line breaks with \\

\author{A.~Thornthwaite}
\affiliation{Oliver Lodge Laboratory, University of Liverpool,
Liverpool, L69 7ZE, United Kingdom.}

\author{D.~O'Donnell}
%\thanks{}
\email[Corresponding author:]{ david.odonnell@liv.ac.uk}
\affiliation{Oliver Lodge Laboratory, University of Liverpool,
Liverpool, L69 7ZE, United Kingdom.}

\author{R.D.~Page}
\affiliation{Oliver Lodge Laboratory, University of Liverpool,
Liverpool, L69 7ZE, United Kingdom.}

\author{D.T.~Joss}
\affiliation{Oliver Lodge Laboratory, University of Liverpool,
Liverpool, L69 7ZE, United Kingdom.}

\author{C.~Scholey}
\affiliation{Department of Physics, University of Jyv\"askyl\"a,
PO Box 35, FI-40014, Jyv\"askyl\"a, Finland.}

\author{L.~Bianco}
\altaffiliation[Present address:]{DESY, Notkestra{\ss}e 85, 22607 Hamburg, Germany}
\affiliation{Department of Physics, University of Guelph,
Guelph, Ontario, N1G 2W1, Canada.}

\author{L.~Capponi}
\affiliation{School of Engineering, University of the West of Scotland,
High Street, Paisley, PA1 2BE, United Kingdom.}

\author{R.J.~Carroll}
\affiliation{Oliver Lodge Laboratory, University of Liverpool,
Liverpool, L69 7ZE, United Kingdom.}

\author{I.G.~Darby}
\altaffiliation[Present address: ]{IAEA Nuclear Spectrometry and Applications Laboratory, Physics Section, A-2444, Siebersdorf, Austria.}
\affiliation{Department of Physics, University of Jyv\"askyl\"a, PO Box 35, FI-40014, Jyv\"askyl\"a, Finland.}

\author{L.~Donosa}
\affiliation{Oliver Lodge Laboratory, University of Liverpool,
Liverpool, L69 7ZE, United Kingdom.}

\author{M.C.~Drummond}
\affiliation{Oliver Lodge Laboratory, University of Liverpool,
Liverpool, L69 7ZE, United Kingdom.}

\author{F.~Ertu\u{g}ral}
\affiliation{Physics Department, Faculty of Arts and Sciences,
Sakarya University, 54100 Serdivan, Adapazari, Turkey.}

\author{T.~Grahn}
\affiliation{Department of Physics, University of Jyv\"askyl\"a,
PO Box 35, FI-40014, Jyv\"askyl\"a, Finland.}

\author{P.T.~Greenlees}
\affiliation{Department of Physics, University of Jyv\"askyl\"a,
PO Box 35, FI-40014, Jyv\"askyl\"a, Finland.}

\author{K.~Hauschild}
\affiliation{Department of Physics, University of Jyv\"askyl\"a,
PO Box 35, FI-40014, Jyv\"askyl\"a, Finland.}
\affiliation{CSNSM, IN2P3-CNRS et Universit\'{e} Paris Sud,
Paris, France}

\author{A.~Herzan}
\affiliation{Department of Physics, University of Jyv\"askyl\"a,
PO Box 35, FI-40014, Jyv\"askyl\"a, Finland.}

\author{U.~Jakobsson}
\affiliation{Department of Physics, University of Jyv\"askyl\"a,
PO Box 35, FI-40014, Jyv\"askyl\"a, Finland.}

\author{P.~Jones}
\affiliation{Department of Physics, University of Jyv\"askyl\"a,
PO Box 35, FI-40014, Jyv\"askyl\"a, Finland.}

\author{R.~Julin}
\affiliation{Department of Physics, University of Jyv\"askyl\"a,
PO Box 35, FI-40014, Jyv\"askyl\"a, Finland.}

\author{S.~Juutinen}
\affiliation{Department of Physics, University of Jyv\"askyl\"a,
PO Box 35, FI-40014, Jyv\"askyl\"a, Finland.}

\author{S.~Ketelhut}
\affiliation{Department of Physics, University of Jyv\"askyl\"a,
PO Box 35, FI-40014, Jyv\"askyl\"a, Finland.}

\author{M.~Labiche}
\affiliation{STFC Daresbury Laboratory, Daresbury, Warrington,
WA4 4AD, United Kingdom.}

\author{M.~Leino}
\affiliation{Department of Physics, University of Jyv\"askyl\"a,
PO Box 35, FI-40014, Jyv\"askyl\"a, Finland.}

\author{A.~Lopez-Martens}
\affiliation{Department of Physics, University of Jyv\"askyl\"a,
PO Box 35, FI-40014, Jyv\"askyl\"a, Finland.}
\affiliation{CSNSM, IN2P3-CNRS et Universit\'{e} Paris Sud,
Paris, France}

\author{K.~Mullholland}
\affiliation{School of Engineering, University of the West of Scotland,
High Street, Paisley, PA1 2BE, United Kingdom.}

\author{P.~Nieminen}
\affiliation{Department of Physics, University of Jyv\"askyl\"a,
PO Box 35, FI-40014, Jyv\"askyl\"a, Finland.}

\author{P.~Peura}
\affiliation{Department of Physics, University of Jyv\"askyl\"a,
PO Box 35, FI-40014, Jyv\"askyl\"a, Finland.}

\author{P.~Rahkila}
\affiliation{Department of Physics, University of Jyv\"askyl\"a,
PO Box 35, FI-40014, Jyv\"askyl\"a, Finland.}

\author{S.~Rinta-Antila}
\affiliation{Department of Physics, University of Jyv\"askyl\"a,
PO Box 35, FI-40014, Jyv\"askyl\"a, Finland.}

\author{P.~Ruotsalainen}
\affiliation{Department of Physics, University of Jyv\"askyl\"a,
PO Box 35, FI-40014, Jyv\"askyl\"a, Finland.}

\author{M.~Sandzelius}
\affiliation{Department of Physics, University of Jyv\"askyl\"a,
PO Box 35, FI-40014, Jyv\"askyl\"a, Finland.}

\author{J.~Sar\'en}
\affiliation{Department of Physics, University of Jyv\"askyl\"a,
PO Box 35, FI-40014, Jyv\"askyl\"a, Finland.}

\author{B.~Say\u{g}i}
\affiliation{Oliver Lodge Laboratory, University of Liverpool,
Liverpool, L69 7ZE, United Kingdom.}

\author{J.~Simpson}
\affiliation{STFC Daresbury Laboratory, Daresbury, Warrington,
WA4 4AD, United Kingdom.}

\author{J.~Sorri}
\affiliation{Department of Physics, University of Jyv\"askyl\"a,
PO Box 35, FI-40014, Jyv\"askyl\"a, Finland.}

\author{J.~Uusitalo}
\affiliation{Department of Physics, University of Jyv\"askyl\"a,
PO Box 35, FI-40014, Jyv\"askyl\"a, Finland.}

\date{\today}% It is always \today, today,

\begin{abstract}
The $\alpha$-decay chains originating from the $\pi s_{1/2}$ and $\pi h_{11/2}$ states in $^{173}$Au have been investigated following fusion-evaporation reactions. Four generations of $\alpha$ radioactivities have been correlated with $^{173}$Au$^m$ leading to a measurement of the $\alpha$ decay of $^{161}$Ta$^m$. It has been found that the known $\alpha$ decay of $^{161}$Ta, which was previously associated with the decay of the ground state, is in fact the decay of an isomeric state. This work also reports on the first observation of prompt $\gamma$ rays feeding the ground state of $^{173}$Au. This prompt $\gamma$ radiation was used to aid the study of the $\alpha$-decay chain originating from the $\pi s_{1/2}$ state in $^{173}$Au. Three generations of $\alpha$ decays have
been correlated with this state leading to the observation of a previously unreported activity which is assigned as the decay of $^{165}$Re$^g$. This work also reports the excitation energy of an $\alpha$-decaying isomer in $^{161}$Ta and the $Q_\alpha$-value of the decay of $^{161}$Ta$^g$.

\end{abstract}

\maketitle

\section{Introduction}
Proton and $\alpha$-decay $Q$-value measurements provide important information on the nuclear mass surface far from the valley of $\beta$ stability.
There are many examples of long chains of $\alpha$ decays between nuclear ground states which, if connected to a nuclide of known mass excess, would
allow the mass excesses of nuclei beyond the proton drip line to be determined. Often these nuclei are too short-lived or too weakly produced to be
measured directly by precision methods such as Schottky mass spectrometry or Penning trap mass spectrometry.

\begin{figure}
\centering
\includegraphics[width=0.46\textwidth,angle=0,clip]{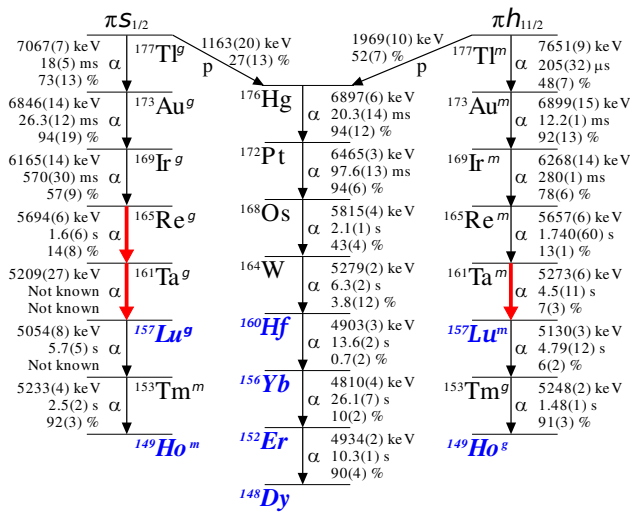}
\caption{(Color online) Schematic decay chains originating from the $\pi s_{1/2}$ ground state of $^{177}$Tl and the $\pi h_{11/2}$ isomeric state in $^{177}$Tl. The nuclides written in blue bold italic font have had their mass excesses measured directly~\cite{Beck_masses,Litvinov_masses}. The decay Q-values, half-lives and branching ratios are indicated. Data not measured as a result of the present study are taken from Refs.~\cite{Poli,Kettunen,Page_alpha,Rytz,Hofmann81,Hofmann79,A161,A153,A176,A168,A164,A160,A156,A152,A149,A148}. Those decays which have been observed or deduced for the first time or have been reassigned as a result of the present work are indicated by wider red arrows.}
\label{chains}
\end{figure}

A representative example of direct relevance to the present study is the work of Poli {\it et al.}~\cite{Poli}, who measured proton and $\alpha$-particle
emission from the $\pi s_{1/2}$ ground state and $\pi h_{11/2}$ isomeric state in $^{177}$Tl. As illustrated in Fig.~\ref{chains}, the daughter of
$^{177}$Tl proton decays is $^{176}$Hg. The $\alpha$-decay $Q$-values of the decay chain of $^{176}$Hg are known down to $^{152}$Er, which decays to
$^{148}$Dy the mass of which is known from Penning trap measurements~\cite{Beck_masses}. The experimental data therefore allowed the mass excesses of the $\pi
s_{1/2}$ ground states and $\pi h_{11/2}$ isomeric states of $^{177}$Tl, $^{173}$Au, $^{169}$Ir and $^{165}$Re to be deduced. The mass excesses of
$^{152}$Er, $^{156}$Yb and $^{160}$Hf were subsequently measured directly using Schottky mass spectrometry~\cite{Litvinov_masses} and the values obtained
were consistent with those deduced from the mass excess of $^{148}$Dy and the relevant $\alpha$-decay $Q$-values.

In the present work, the known $\alpha$ decay of $^{161}$Ta is shown to be correlated with the $\alpha$ decays of the $\pi h_{11/2}$ states in
$^{173}$Au, $^{169}$Ir and $^{165}$Re, indicating that it also originates from a $\pi h_{11/2}$-based state, rather than a $\pi s_{1/2}$-based state as had been
previously assumed~\cite{A157}. This establishes the complete $\alpha$-decay chain from the $\pi h_{11/2}$ isomeric state in the proton-unbound
$^{177}$Tl down to the corresponding state in $^{149}$Ho, which constitutes the ground state of this nuclide~\cite{Toth_153Tm}. In addition, a previously
unobserved radioactivity has been correlated with the $\alpha$ decay of the ground states of $^{173}$Au and $^{169}$Ir indicating that this new decay originates from the $\pi s_{1/2}$-based ground state of $^{165}$Re. As the masses of both $^{148}$Dy~\cite{Beck_masses} and $^{149}$Ho~\cite{Litvinov_masses} have
been measured, this study not only provides a cross check of the mass excesses deduced in Ref.~\cite{Poli} but, perhaps more importantly, allows the
$Q$-value of the $\alpha$ decay of the $\pi s_{1/2}$-based state in $^{161}$Ta and the energy difference between this state and the $\pi h_{11/2}$-based state to be deduced.

\section{Experimental details}

The $^{173}$Au nuclei were produced via fusion-evaporation reactions induced by the bombardment of a 0.5~mg/cm$^{2}$ $^{92}$Mo target of 97~\%
isotopic enrichment with a beam of $^{84}$Sr$^{16+}$ ions. The Sr beam was provided by the K130 cyclotron of the Accelerator Laboratory of the
University of Jyv\"{a}skyl\"{a} with an energy of 392~MeV for approximately 140 hours and 400~MeV for approximately 145 hours with an average beam
current of 150~enA. The change in beam energy was performed with an aim to increase the production of $^{173}$Hg~\cite{ODonnell_173Hg} nuclei which was the primary aim of the experiment.

At the target position, 34 high-purity Ge detectors, ten of EUROGAM Phase 1 type~\cite{Beausang_Jurogam} and 24 clover detectors~\cite{Duchene_clover}, were positioned to facilitate the detection of prompt $\gamma$ radiation. The recoiling fusion-evaporation residues (recoils) were separated from the unreacted beam using the RITU He-filled magnetic separator~\cite{Leino2} and were transported to the RITU focal plane where the GREAT spectrometer~\cite{Page_GREAT} was located. Here the recoils traversed an isobutane-filled multi-wire proportional chamber (MWPC) before implanting into one of two 300~$\mu$m-thick double-sided Si strip detectors (DSSD). The average rate of recoil implantations was found to be $\approx$ 150~Hz across both of the DSSDs. In addition to the MWPC and the DSSDs, the GREAT spectrometer comprised a box of 28 Si PIN diode detectors, a planar Ge detector and four clover-type Ge detectors. The MWPC provided energy loss and (in conjunction with the DSSDs) time-of-flight information, which was used to separate the recoils from any residual scattered beam. The signals from all detectors were passed to the Total Data Readout aquisition system~\cite{TDR} where they were time stamped with a precision of 10~ns to facilitate temporal correlations between the implantation of recoils in the DSSDs and their subsequent radioactive decays. The data were analysed using the Grain software package~\cite{Rahkila_Grain}.

\section{Experimental results}

\subsection{Alpha decay of $^{173}$Au$^m$}
There are considerable experimental challenges in establishing decay correlations with low-$Z$ members of $\alpha$-decay chains. Firstly, the
relatively long half-lives mean that correlations can become obscured by interfering radioactivities produced as a result of competing reaction channels. Secondly, low $\alpha$-decay branching ratios mean that large samples of parent nuclei are typically needed to facilitate such studies.
However, these are generally produced with rather lower cross sections than their descendents thus further compounding the difficulties of studying long $\alpha$-decay chains.

\begin{figure}
\centering
\includegraphics[width=0.46\textwidth,angle=0,clip]{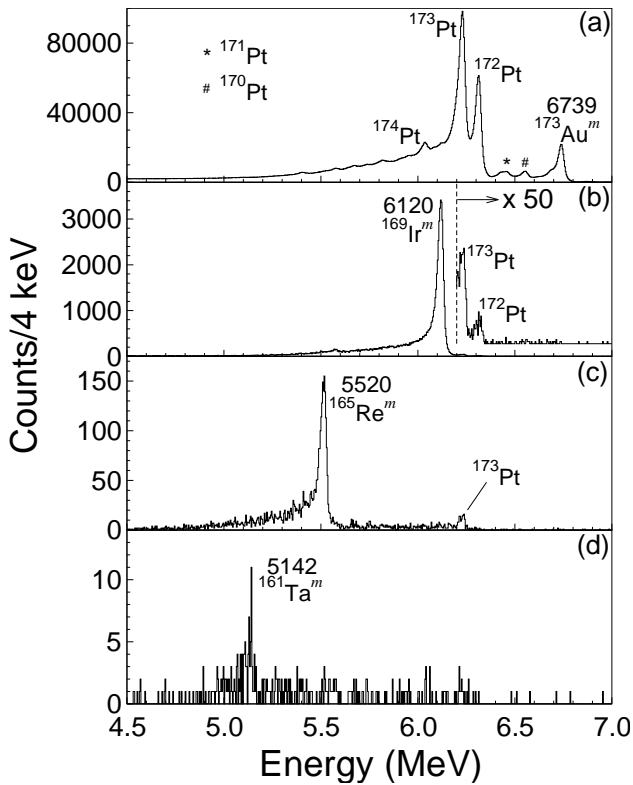}
\caption{Spectra showing (a) all decays observed in the DSSDs within 100~ms of a recoil implantation in the same pixel; (b) all decays following the
detection of a $^{173}$Au$^m$ $\alpha$ decay occurring within 100~ms of a recoil implantation in the same pixel; (c) all decays following the
observation of both $^{173}$Au$^m$ and $^{169}$Ir$^m$ $\alpha$ decays such that the former were detected within 100~ms of a recoil implantation in the same
pixel; (d) all fourth-generation $\alpha$ decays following the observation of an initial $\alpha$ decay within 100~ms of recoil implantation and the subsequent detection of the $\alpha$ decays of both $^{169}$Ir$^m$ and $^{165}$Re$^m$.}
\label{isomer_alphas}
\end{figure}

In the present study, approximately 190,000 recoil-$\alpha$($^{173}$Au$^m$) events were produced in order to provide a sufficient sample of correlated
$^{169}$Ir$^m$, $^{165}$Re$^m$ and $^{161}$Ta$^m$ $\alpha$ decays. The decay chain was analysed by selecting $^{173}$Au$^m$ $\alpha$ decays occurring within 100~ms of a recoil being implanted into the same DSSD pixel. All decays fulfilling this criterion are shown in Fig.~\ref{isomer_alphas}(a).
The only other conditions imposed on the search for subsequent decays were that they all had to have been observed in the same pixel and within 15~s
of the implantation. The energy of the $\alpha$ particles corresponding to the decay of $^{173}$Au$^m$ was found to be 6739(15)~keV which is in good agreement with the error-weighted mean of previous measurements of this activity~\cite{A169}. It is worth noting at this point that the DSSDs were calibrated using $\alpha$ lines at 5407, 6038 and 6315~keV corresponding to the known activities of $^{170}$Os~\cite{A166} (not visible in Fig.~\ref{isomer_alphas}(a)), $^{174}$Pt~\cite{A170} and $^{172}$Pt~\cite{A168}, respectively. In addition to the energy measurement the difference in time between each recoil implantation and the detection of $^{173}$Au$^m$ $\alpha$ decays was recorded and the distribution was fitted using the least-squares method. The half-life of this radioactivity was measured to be 12.2(1)~ms which is consistent with previous measurements~\cite{A169}.

The first decays observed following the $^{173}$Au$^m$ $\alpha$ decays are shown in Fig.~\ref{isomer_alphas}(b). The dominant peak in this spectrum
has an energy of 6120(14)~keV which is consistent with previous measurements of the $\alpha$ decay of an isomeric state in $^{169}$Ir. The half-life of this decay was measured using a similar method to the $^{173}$Au$^m$ activity except in this case the difference in time between the $^{173}$Au$^m$ and $^{169}$Ir$^m$ $\alpha$ decays were recorded. A least-squares fit to this distribution yielded a half-life of 280(1)~ms which is in good agreement with previous studies~\cite{Poli}. A branching ratio of 78(6)~\% was measured for the $\alpha$ decay of this state which agrees well with the value of 84(8)~\% reported by Poli {\it et al.}~\cite{Poli}. In addition to the main peak, a further two, relatively weak, peaks can be seen in Fig.~\ref{isomer_alphas}(b) which have been identified as resulting from the $\alpha$ decay of the ground states of $^{173}$Pt and $^{172}$Pt. These nuclei represent the most prominent $\alpha$ decays observed in this study and their presence in the spectra of Fig.~\ref{isomer_alphas}(b) can be understood as a consequence of the incorrect correlation of $^{173}$Au$^m$ $\alpha$ decays with implanted $^{173}$Pt and $^{172}$Pt nuclei. This also explains the appearance of a peak associated with the $\alpha$ decay of $^{173}$Pt in Fig.s~\ref{isomer_alphas}(c) and (d).

Shown in Fig.~\ref{isomer_alphas}(c) are those decays observed following the detection of a $^{173}$Au$^m$ $\alpha$ decay occurring within 100~ms of
an implanted recoil and the subsequent detection of an $^{169}$Ir$^m$ $\alpha$ decay, all in the same pixel. The dominant peak in this spectrum has a
measured energy of 5520(6)~keV which is in good agreement with the previous observations of the decay of $^{165}$Re$^m$~\cite{Page_alpha}. A half-life of 1740(60)~ms was measured for this activity which is in good agreement with earlier work~\cite{Page_alpha}. In addition, a branching ratio of 13(1)~\% was measured which agrees well with the value of 13(3)~\% established in earlier work~\cite{A161}.

Fig.~\ref{isomer_alphas}(d) shows all decays which were preceded by an $\alpha$ decay within 100~ms of the implantation of a recoil followed by the
sequential observation of the $\alpha$ decays of $^{169}$Ir$^m$ and $^{165}$Re$^m$. The dominant peak in this spectrum has an energy of 5142(6)~keV. The
energy of this activity is in good agreement with the previously reported value for the $\alpha$ decay of $^{161}$Ta (5148(5)~keV~\cite{A157}). The maximum-likelihood method~\cite{Schmidt_max_likelihood} was used to fit the time distribution between the detection of $^{165}$Re$^m$ and $^{161}$Ta $\alpha$ decays and has yielded a half-life of 4.5(11)~s. This value is consistent with previous measurements~\cite{Page_alpha}. The branching ratio for the $\alpha$ decay of this state was found to be 7(3)~\%. This represents the first occasion that this ratio has been reported as a result of an experimental measurement. The measured value is in good agreement with the estimated branching ratio of 5~\%~\cite{A157}.

\subsection{Alpha decay of $^{173}$Au$^g$}
\begin{figure}
\centering
\includegraphics[width=0.46\textwidth,angle=0,clip]{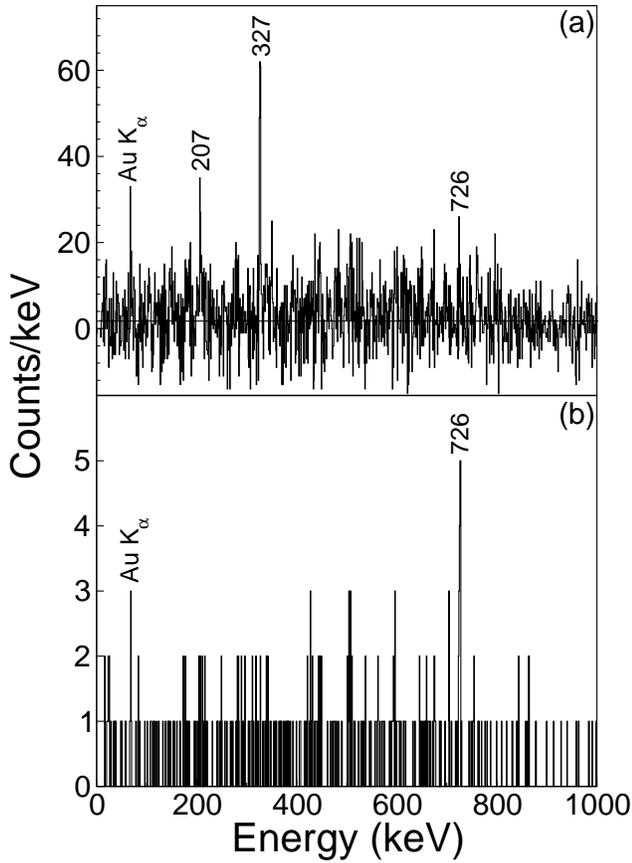}
\caption{(a) Background-subtracted prompt $\gamma$-ray spectrum in delayed coincidence with $^{173}$Au$^g$ $\alpha$ decay. (b) Projection of a
$^{173}$Au$^g$ $\alpha$-tagged $\gamma\gamma$ matrix gated on a 327~keV transition.}
\label{gammas}
\end{figure}

Previous measurements of the energy of the $\alpha$ decay of the ground state of $^{173}$Au have yielded an error-weighted mean value of
6683(9)~keV~\cite{A169}. The proximity of this line to that corresponding to the decay of the $^{173}$Au isomeric state means that isolating the
$\alpha$-decay chain associated with the $^{173}$Au ground state is problematic. To aid the analysis, prompt $\gamma$ rays
feeding the $^{173}$Au ground state were investigated. Fig.~\ref{gammas}(a) shows a background-subtracted spectrum of prompt $\gamma$ rays observed
at the target position in delayed coincidence with $\alpha$ particles having an energy consistent with the decay of the ground state of $^{173}$Au. The $\gamma$-recoil-$\alpha$($^{173}$Au$^g$) events were only considered if the $\alpha$ particle was detected within 150~ms and in the same pixel as the implanted recoil. Three $\gamma$ rays, of energies 207, 327 and 726~keV, have been identified for the first time as feeding the ground state of $^{173}$Au. A $^{173}$Au$^g$
$\alpha$-tagged $\gamma\gamma$ matrix was also constructed and Fig.~\ref{gammas}(b) shows those $\gamma$ rays observed in coincidence with the
327~keV transition. Although this spectrum demonstrates that the 327 and 726~keV transitions are mutually coincident, it has not been possible on the basis of the present data to construct a level scheme.

\begin{figure}
\centering
\includegraphics[width=0.46\textwidth,angle=0,clip]{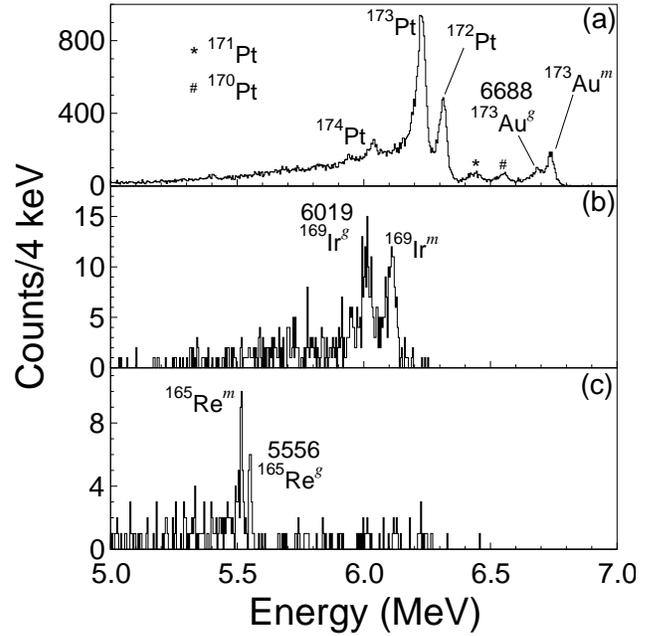}
\caption{Spectra showing (a) all decays observed in the DSSDs within 100~ms of a recoil implantation in the same pixel and in delayed coincidence with prompt $\gamma$ rays of 327 or 726~keV; (b) all second-generation decays following the detection of a $^{173}$Au$^g$ $\alpha$ decay occurring within 100~ms of a recoil implantation in the same pixel and with the same $\gamma$-ray conditions as (a); (c) third-generation decays, subject to the same conditions as (b) but which were also preceded by an $\alpha$ decay having an energy $< 6030$~keV.}
\label{ground_alphas}
\end{figure}

Fig.~\ref{ground_alphas}(a) shows decays occurring within 100~ms of the implantation of a recoil with the added condition that they were also in delayed coincidence with either a 327 or 726~keV $\gamma$ ray detected at the target position. The relatively low $\gamma$-ray detection efficiency ($\sim 5~\%$) at the target position, combined with the background introduced by Compton scattering, ensures that this gating technique is not sufficiently selective to isolate the $^{173}$Au$^g$ $\alpha$ decays but has the effect of enhancing them (compare Figs.~\ref{isomer_alphas}(a) and \ref{ground_alphas}(a)). The $^{173}$Au$^g$ $\alpha$ decays have been measured to have an energy of 6688(14)~keV and using the least-squares method the half-life of the activity was found to be 26.3(12)~ms. Both the energy and half-life values measured here agree well with previous measurements~\cite{Poli}.

Fig.~\ref{ground_alphas}(b) shows those decays observed to follow $^{173}$Au$^g$ $\alpha$ decays which occurred in the same pixel and within 100~ms of the implantation of a recoil. The two peaks in this spectrum correspond to $\alpha$ decays with energies of 6019(14) and 6120(14)~keV. These values are consistent with the previously reported energies for the decay of the ground state and an isomeric state of $^{169}$Ir, respectively. A low-energy tail on the $^{173}$Au$^m$ peak, arising from a combination of escaping $\alpha$ particles and the effects of radiation damage on the Si detectors, results in a spectrum which contains both $^{169}$Ir$^g$ and $^{169}$Ir$^m$ $\alpha$ decays. The number of $^{169}$Ir$^m$ $\alpha$ decays present in Fig.~\ref{ground_alphas}(b) is consistent with the number of $^{173}$Au$^m$ $\alpha$ decays included in the gate used to identify the $\alpha$ decay of the ground state of $^{173}$Au.

The half-life of the radioactivity associated with the decay of the $^{173}$Au ground state was measured to be 570(30)~ms is in good agreement with previous studies~\cite{Poli}. The branching ratio for the $\alpha$ decay of this state was found to be 57(9)~\%, which is in good agreement with the value of 45(15)~\% obtained in previous studies~\cite{A165}.

Third-generation $\alpha$ decays which follow the decay of both $^{173}$Au$^g$ and $^{169}$Ir$^g$ and are in delayed coincidence with 327 or 726~keV prompt $\gamma$ rays, are shown in Fig.~\ref{ground_alphas}(c). The two peaks visible in this spectrum correspond to $\alpha$ decays of energies 5520(6) and 5556(6)~keV. The former is in good agreement with the previously reported energies of the decay of $^{165}$Re$^m$. Similarly to the situation discussed above, the number of $^{165}$Re$^m$ $\alpha$ decays observed in Fig.~\ref{ground_alphas}(c) is consistent with the number of $^{169}$Ir$^m$ included in the gate used to identify the ground state decays of $^{169}$Ir. The second peak is a previously unreported acitivity. The clear correlation of this new activity with the decays of both $^{173}$Au$^g$ and $^{169}$Ir$^g$ leads to its assignment as the $\alpha$ decay of the ground state of $^{165}$Re. To determine the half-life of this new activity, the $\alpha(^{169}$Ir$^g$)-$\alpha(^{165}$Re$^g$) time spectrum was fitted using the maximum-likelihood method~\cite{Schmidt_max_likelihood} and this yielded a value of 1.6(6)~s. A branching ratio of 14(8)~\% has been measured for the $\alpha$ decay of this state.

\section{Discussion}
\begin{figure}
\centering
\includegraphics[width=0.46\textwidth,angle=0,clip]{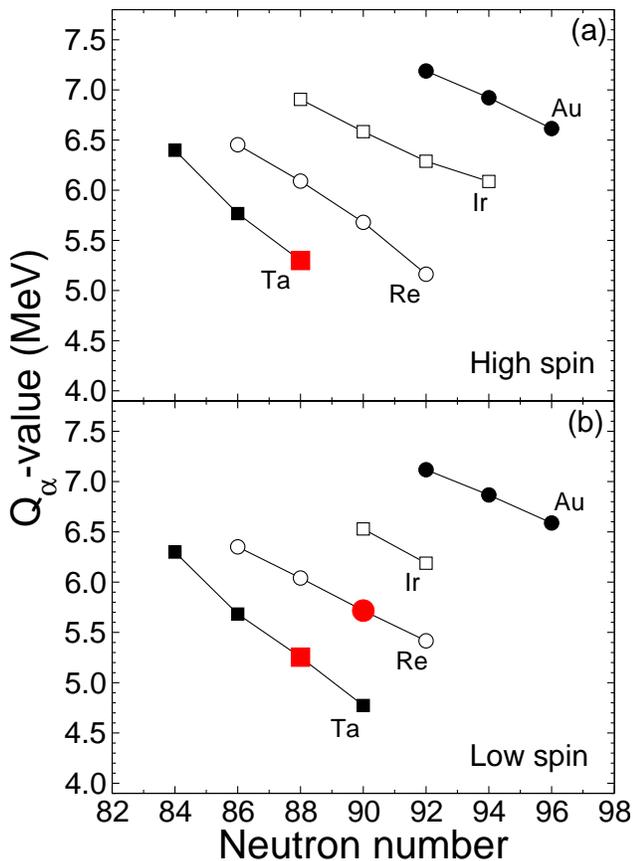}
\caption{(Color online) Experimental $\alpha$-decay $Q$ values of odd-$Z$, even-$N$ nuclei relevant to the discussion of the present work. In panel (a) is shown the $Q_\alpha$ values for high-spin states (associated with $\pi h_{11/2}$ orbital) while in (b) the $Q_\alpha$ values for the low-spin states ($\pi s_{1/2}$) are shown. The symbols representing the $Q_\alpha$ values determined as a result of the present work have been colored red and enlarged. Data not measured as a result of the present work are taken from Refs.~\cite {Kettunen,A171,Rowe,Davids97,A163,Rytz,A157,A159,Meissner92,A153,Irvine97,A155}.}
\label{Q-values}
\end{figure}

Prior to the undertaking of this work, the known $\alpha$ decay of $^{161}$Ta was assumed to be the result of the decay of the ground state~\cite{A157}. However, from the results presented above it is apparent that the 5142(6)~keV $\alpha$ decay is in fact the decay of the high-spin isomeric state in $^{161}$Ta. The deduced $Q_\alpha$ value of 5273(6)~keV is plotted in Fig.~\ref{Q-values}(a) and can be seen to continue the near-linear trend of the decreasing $Q_\alpha$ values with increasing neutron number.

The $Q_\alpha$-value for the previously unreported decay of the ground state of $^{165}$Re is plotted in Fig.~\ref{Q-values}(b). This value appears to fit very well with the linear trend already established by the neighboring Re isotopes. In combining this new measurement with the $\alpha$-decay $Q$ values of Fig.~\ref{chains} and the mass excess of $^{156}$Yb reported by Litvinov {\it et al.} (-~53283(28)~keV~\cite{Litvinov_masses}), it is possible to determine the mass excess for the ground state of $^{161}$Ta: -~38816(40)~keV.

As a result of the observation of the $\alpha$ decay of $^{165}$Re$^g$ in the present work, it is possible to determine the excitation energy of the high-spin state in $^{161}$Ta which is given by:
\begin{equation}
\Delta E (^{161}\text{Ta}) = \Delta E (^{165}\text{Re}) + Q_\alpha(^{165}\text{Re}^g) - Q_\alpha(^{165}\text{Re}^m). \nonumber \\
\end{equation} Using the $\alpha$-decay energies reported here and the excitation energy of the $\alpha$-decaying isomeric state in $^{173}$Au, 214(23)~keV as reported by Poli {\it et al.}~\cite{Poli}, it has been determined that the high-spin state in $^{161}$Ta has an excitation energy of 95(38)~keV. Taking this analysis one step further, the knowledge of the energy difference of the two $\alpha$-decaying states in $^{157}$Lu, 26(7)~keV~\cite{A157}, allows the $Q_\alpha$-value of the unobserved decay of the ground state of $^{161}$Ta to be determined: $Q_\alpha$($^{161}$Ta$^g$) = 5204(39)~keV. This new value is plotted in Fig.~\ref{Q-values}(b) where once more it fits well with the trend established by the neighboring Ta isotopes.

Using the measured mass excess for $^{156}$Yb and the $\alpha$-decay $Q$ values of Fig.~\ref{chains}, the mass excesses of the ground and isomeric states in $^{149}$Ho can be deduced. The deduced mass excess of the high-spin state in $^{149}$Ho is -61648(40)~keV which agrees remarkably well with the directly measured value of Litvinov {\it et al.}~\cite{Litvinov_masses} of -61646(31)~keV. The mass excess deduced for the low-spin state of $^{149}$Ho was found to be -61582(58)~keV which is in-line with expectations based on the previously known 49~keV excitation energy of the $\pi s_{1/2}$-based isomer in $^{149}$Ho~\cite{A149}.

\begin{figure}
\centering
\includegraphics[width=0.34\textwidth,angle=270,clip]{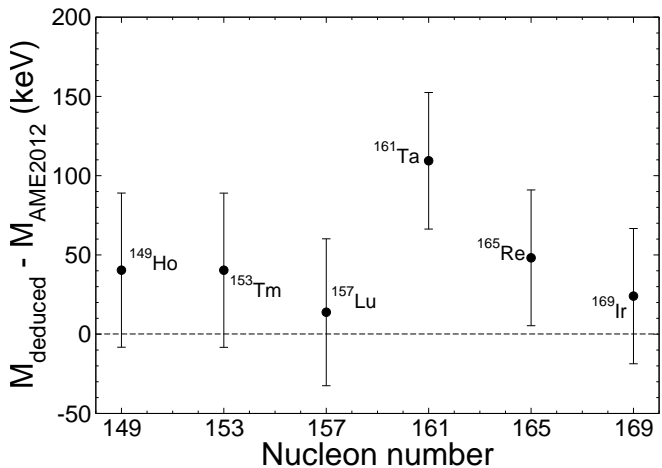}
\caption{Differences between mass excesses deduced in the present work and values reported in the most recent Atomic Mass Evaluation~\cite{AME2012_1,AME2012_2}. The dashed line indicates zero difference.}
\label{masses}
\end{figure}

The mass excesses deduced in the present work are compared with the values reported in the most recent Atomic Mass Evaluation (AME2012)~\cite{AME2012_1,AME2012_2} in Fig.~\ref{masses}. Overall, there is very good agreement between the values obtained in this study and those in the evaluation with the deduced mass of $^{161}$Ta being the notable exception. This discrepancy is possibly the result of the inclusion of the incorrectly assigned $\alpha$-decay of $^{161}$Ta. Indeed, if the 69~keV energy difference between the $Q_\alpha$ values of $^{161}$Ta$^m$ and $^{161}$Ta$^g$ is taken into account then the difference between the mass reported here and the AME2012 value is similar to those found for the other five nuclides plotted in Fig.~\ref{masses}.

The consistency in the mass measurements indicated by the agreement between the deduced masses of the ground and isomeric states of $^{149}$Ho and the masses measured in Ref.~\cite{Litvinov_masses} suggests that all of the $\alpha$ decays proceed between ground states with no electromagnetic decays occurring at any points in the decay chain between $^{177}$Tl and $^{149}$Ho. This is indicative that the single-particle configurations, established as $\pi h_{11/2}$ and $\pi s_{1/2}$ in the heavier members of the decay chain, are also consistent down the entire decay chain. This conclusion is supported by the reduced width measurements, calculated using the Rasmussen formalism~\cite{Rasmussen} and assuming $s$-wave emisssion, which are listed in Table~\ref{reduced_widths}. The reduced widths measured in the present work have been compared to the value corresponding to the $\alpha$ decay of the ground state of $^{212}$Po. These hindrance factors, also listed in Table~\ref{reduced_widths}, are consistent with unhindered $\alpha$ decays.

\begin{table}[!th]% add [H] placement to break table across pages
\caption{Reduced widths, $\delta^2$, and hindrance factors, HF, for the $\alpha$ decay of nuclei measured in the present work. The hindrance factors have been measured relative to the ground state to ground state $\alpha$ decay of $^{212}$Po.}
\begin{ruledtabular}
\begin{tabular}{ccc}
\bf & $\delta^2$ (keV) & HF\\ \hline \\
$^{169}$Ir$^m$ & 70(10) & 1.00(15)\\
$^{169}$Ir$^g$ & 64(13) & 0.91(19)\\
$^{165}$Re$^m$ & 81(8) & 1.15(12)\\
$^{165}$Re$^g$ & 66(45) & 0.93(64)\\
$^{161}$Ta$^m$ & 113(56) & 1.60(80)\\
\end{tabular}
\label{reduced_widths}
\end{ruledtabular}
\end{table}

In Ref.~\cite{Lagergren_161Ta} an extensive level scheme of excited states in $^{161}$Ta built upon a proposed $J^\pi = 11/2^-$ state was reported. However, in that work it was not possible to establish whether this level or a $9/2^-$ level was the lowest-lying $\pi h_{11/2}$ state. The separation energy of the $J^\pi = 9/2^-$ and the $11/2^-$ states in the neutron-deficient Ta isotopes is observed to decrease from 99~keV in $^{167}$Ta~\cite{167Ta}, 71~keV in $^{165}$Ta~\cite{165Ta} to 45 keV in $^{163}$Ta~\cite{163Ta}. Extrapolating to $^{161}$Ta suggests the separation could be as low as $\approx$~20 keV in this nuclide. This would be accommodated within the 40~keV uncertainty on the deduced mass excess for the high-spin state in $^{149}$Ho meaning that the question regarding the spin and parity of the $\pi h_{11/2}$-based state in $^{161}$Ta cannot be resolved by the present study. Indeed, it remains unclear whether the $\alpha$-decaying isomer in $^{161}$Ta has $J^\pi = 9/2^-$ or $11/2^-$.

In summary, fusion-evaporation reactions have been used to populate states in $^{173}$Au. Gamma-ray transitions populating the ground state of $^{173}$Au have been identified. In addition, the $\alpha$-decay chains originating from the isomeric $\pi h_{11/2}$ state and the $\pi s_{1/2}$ ground state have been studied, culminating in the observation of the $\alpha$ decay of $^{161}$Ta$^m$ and $^{165}$Re$^g$, respectively. As well as reporting a new activity in the decay of $^{165}$Re$^g$ and confirming that the known $\alpha$ decay of $^{161}$Ta is associated with the high-spin isomer, this work has enabled the relative energies of the $\alpha$-decaying states in $^{161}$Ta to be established. In combining these new measurements with the information already available on $^{157}$Lu it has also been possible to deduce the $Q_\alpha$-value for the decay of the ground state of $^{161}$Ta. As a result of the present work $Q_p$-values of -129(24)~keV and -37(21)~keV have been determined for the ground and isomeric states of $^{161}$Ta, respectively, indicating that these states are only just bound with respect to proton emission.

\begin{acknowledgments}
Financial support for this work has been provided by the UK Science and Technology Facilities Council (STFC) and by the EU 7th framework programme
``Integrating Activities - Transnational Access", project number: 262010(ENSAR) and by the Academy of Finland under the Finnish Centre of Excellence
Programme 2012-2017(Nuclear and Accelerator Based Physics Research at JYFL). TG acknowledges the support of the Academy of Finland, contract number
131665. The authors would like to express their gratitude to the technical staff of the Accelerator Laboratory at the University of Jyv\"askyl\"a for
their support. The authors would also like to thank Charles Reich for stimulating discussions.
\end{acknowledgments}

%\bibliography{daresbury_bibliography}
%\bibliography{MyCollection}

\end{document}